\documentclass[12pt,a4paper]{article}

\pdfoutput=1

\usepackage{graphicx}
\usepackage{subfig}
\usepackage{times}
\usepackage{natbib}
\usepackage{amsmath}
\usepackage[
pdftex=true,
a4paper=true,
draft=false,
debug=false,
colorlinks=true,
linktocpage=true
]
{hyperref}
\usepackage[all]{hypcap}
\hypersetup{
pdfpagemode = {UseOutlines},
baseurl = {http://www.mpia.de/homes/kuiper},
pdftitle = {Radiation pressure feedback in the formation of massive stars},
pdfauthor = {R.~Kuiper},
pdfsubject = {},
pdfkeywords = {}
}

\newlength{\FigureWidth}
\newcommand{\Msol}{\mbox{$\mbox{ M}_\odot$} }

\title{\bf Radiation pressure feedback\\ in the formation of massive stars
}
\author{Rolf Kuiper$^{1,2}$, Hubert Klahr$^2$, Henrik Beuther$^2$, and Thomas Henning$^2$\\
\vspace{1cm}\\
\normalsize $^1$ Argelander Institute for Astronomy, Bonn University, Auf dem H\"ugel 71, D-53121 Bonn, Germany\\ 
\normalsize $^2$ Max Planck Institute for Astronomy, K\"onigstuhl 17, D-69117 Heidelberg, Germany}
\date{\mbox{}}
\begin{document}
\maketitle
\pagestyle{empty}
%
%
\def\bull{\vrule height .9ex width .8ex depth -.1ex}
\makeatletter
\def\ps@plain{\let\@mkboth\gobbletwo
\def\@oddhead{}\def\@oddfoot{\hfil\tiny\bull\quad
``The multi-wavelength view of hot, massive stars''; 39$^{\rm th}$ Li\`ege Int.\ Astroph.\ Coll., 12-16 July 2010 \quad\bull}%
\def\@evenhead{}\let\@evenfoot\@oddfoot}
\makeatother
%
%
\def\beginrefer{\section*{References}%
\begin{quotation}\mbox{}\par}
\def\refer#1\par{{\setlength{\parindent}{-\leftmargin}\indent#1\par}}
\def\endrefer{\end{quotation}}
%
%
{\noindent\small{\bf Abstract:} 
We investigate the radiation pressure feedback in the formation of massive stars in 1, 2, and 3D radiation hydrodynamics simulations of the collapse of massive pre-stellar cores. 
In contrast to previous research, we 
consider frequency dependent stellar radiation feedback, 
resolve the dust sublimation front in the vicinity of the forming star down to 1.27 AU, 
compute the evolution for several $10^5$~yrs covering the whole accretion phase of the forming star, and 
perform a comprehensive survey of the parameter space.
The most fundamental result is that the formation of a massive accretion disk in slowly rotating cores preserves a high anisotropy in the radiation field. 
The thermal radiation escapes through the optically thin atmosphere, effectively diminishing the radiation pressure feedback onto the accretion flow. 
Gravitational torques in the self-gravitating disk drive a sufficiently high accretion rate to overcome the residual radiation pressure. 
Simultaneously, the radiation pressure launches an outflow in the bipolar direction,
which grows in angle with time
and releases a substantial fraction of the initial core mass from the star-disk system.
Summarized, for an initial core mass of 60, 120, 240, and 480\Msol these mechanisms allow the star to grow up to 28.2, 56.5, 92.6, and at least 137.2\Msol respectively.
}

%
%
\section{Introduction}
The understanding of massive stars suffers from the lack of a generally accepted formation scenario.
If the formation of high-mass stars is treated as a scaled-up version of low-mass star formation, a special feature of these high-mass proto-stars is the interaction of the accretion flow with the strong irradiation emitted by the newborn stars due to their short Kelvin-Helmholtz contraction timescale \citep{Shu:1987p1616}. 
Early one-dimensional studies \citep[e.g.][]{Larson:1971p1210, Kahn:1974p1200, Yorke:1977p1358} agree on the fact that the growing radiation pressure potentially stops and reverts the accretion flow onto a massive star.

But this radiative impact strongly depends on the geometry of the stellar environment \citep{Nakano:1989p1267}.
The possibility was suggested to overcome this radiation pressure barrier via the formation of a long-living massive circumstellar disk, which forces the generation of a strong anisotropic feature of the thermal radiation field.
Earlier investigations by \citet{Yorke:2002p1} tried to identify such an anisotropy, but their simulations show an early end of the disk accretion phase shortly after its formation due to strong radiation pressure feedback.
\citet{Krumholz:2009p10975} demonstrated the possibility to maintain the shielding property of the accretion disk by additional feeding due to the so-called `3D radiative Rayleigh-Taylor instability' in the outflow cavity.

\newpage
\section{Method}
\subsection{The aim}
Aim of this study is to reveal the details of the radiation dust interaction in the formation of the most massive star during the collapse of a pre-stellar core (gravitationally unstable fragment of a molecular cloud).
Due to the strong focus onto the core center, the physics in the outer core region have to be simplified, 
e.g.~further fragmentation of the pre-stellar core is suppressed.

\subsection{The code}
\label{sect:Physics}
For this purpose, we use our newly developed self-gravity radiation hydrodynamics code.
The hydrodynamical evolution
is computed using the 
open source code Pluto3 \citep{Mignone:2007p3421}, including full tensor viscosity.
The frequency dependent hybrid radiation transport method is summarized by \citet{Kuiper:2010p12874}.
The implementation of Poisson's equation, the viscosity prescription as well as the tabulated dust and stellar evolution model are presented in 
\citet{Kuiper2010b}.

The simulations are performed on a time independent grid in spherical coordinates.
The radially inner and outer boundary of the computational domain are semi-permeable walls, i.e.~the gas can leave but not enter the computational domain.
The resolution of the non-uniform grid is chosen to be 
$\left(\Delta r \times r~\Delta{\theta} \times r \sin(\theta)~\Delta{\phi}\right)_\mathrm{min} = 1.27 \mbox { AU } \times 1.04 \mbox { AU } \times 1.04 \mbox{ AU}$
in the midplane ($\theta = 90^\circ$) around the forming massive star and decreases logarithmically in the radial outward direction.
The outer core radius is fixed to $r_\mathrm{max} = 0.1$~pc. 
The accurate size of the inner sink cell was determined in a parameter scan presented in 
\citet{Kuiper2010b}, Sect.~5.1.

\subsection{The initial conditions}
\label{sect:InitialConditions}
The basic physical initial conditions are very similar to the one used by \citet{Yorke:2002p1}.
We start from a cold ($T_0 = 20 \mbox{ K}$) pre-stellar core of gas and dust.
The initial dust to gas mass ratio is chosen to be $M_\mathrm{dust} / M_\mathrm{gas} = 1\%$.
The model describes a so-called quiescent collapse scenario without turbulent motion ($\vec{u}_r = \vec{u}_\theta = 0$).
The core is initially in slow solid-body rotation $\left(|\vec{u}_\phi| / R = \Omega_0 = 5\times10^{-13} \mbox{ Hz}\right)$.
The initial density slope drops with $r^{-2}$ and the total mass $M_\mathrm{core}$ varies in the simulations from 60 up to 480 $\mbox{M}_\odot$. 
\begin{figure}[hbt]
\begin{minipage}{0.5\FigureWidth}
\vspace{-3mm}
\includegraphics[width=0.49\FigureWidth]{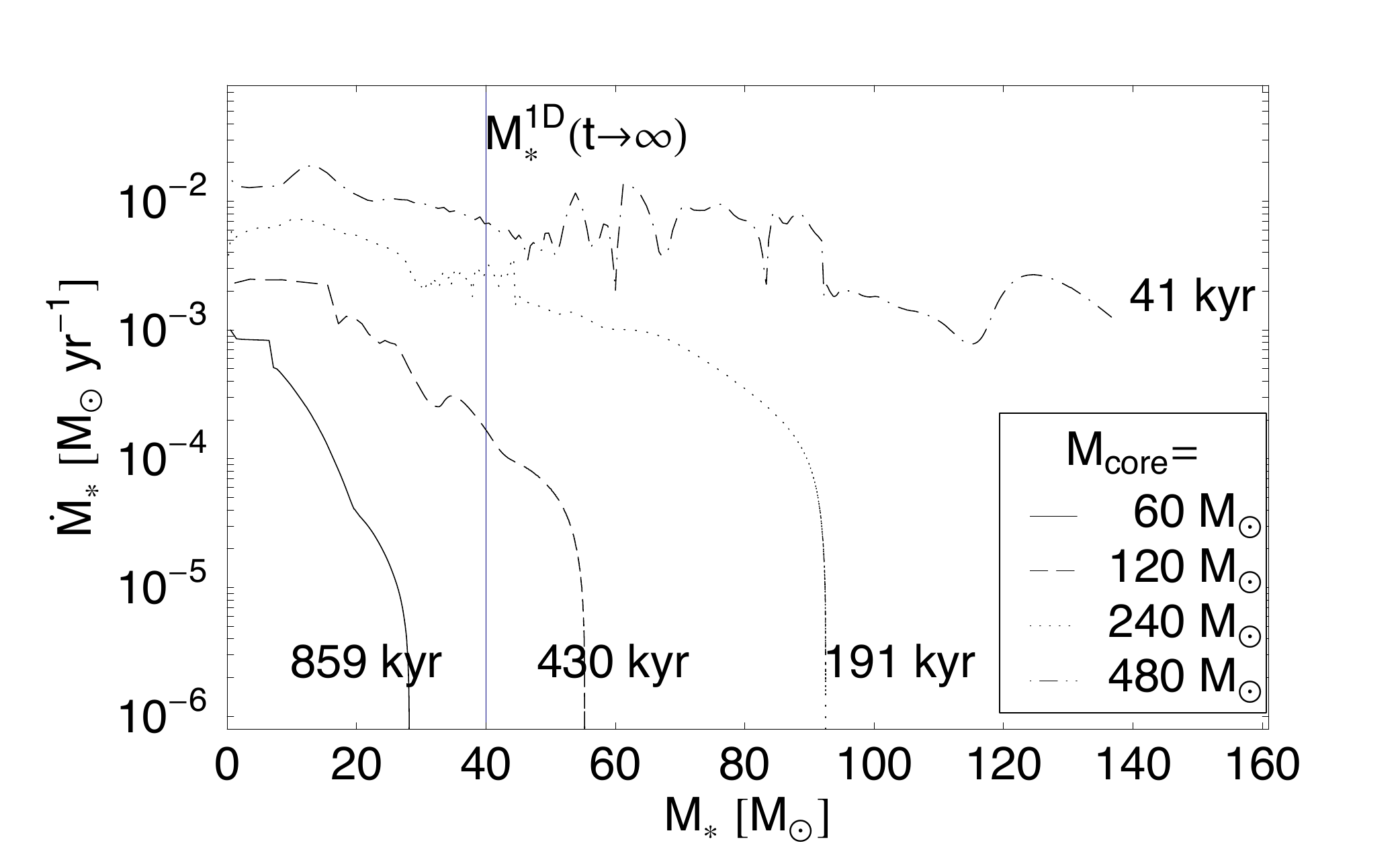}
\caption{
Accretion rate $\dot{M}_*$ as a function of the 
stellar mass $M_*$ for 
varying initial core masses.
The periods of accretion are mentioned for each run.
The vertical line 
marks the upper mass limit for spherically symmetric accretion flows. 
}
\label{fig:2D_McoreScan}
\end{minipage}
\hspace{0.02\FigureWidth}
\begin{minipage}{0.5\FigureWidth}
\includegraphics[width=0.465\FigureWidth]{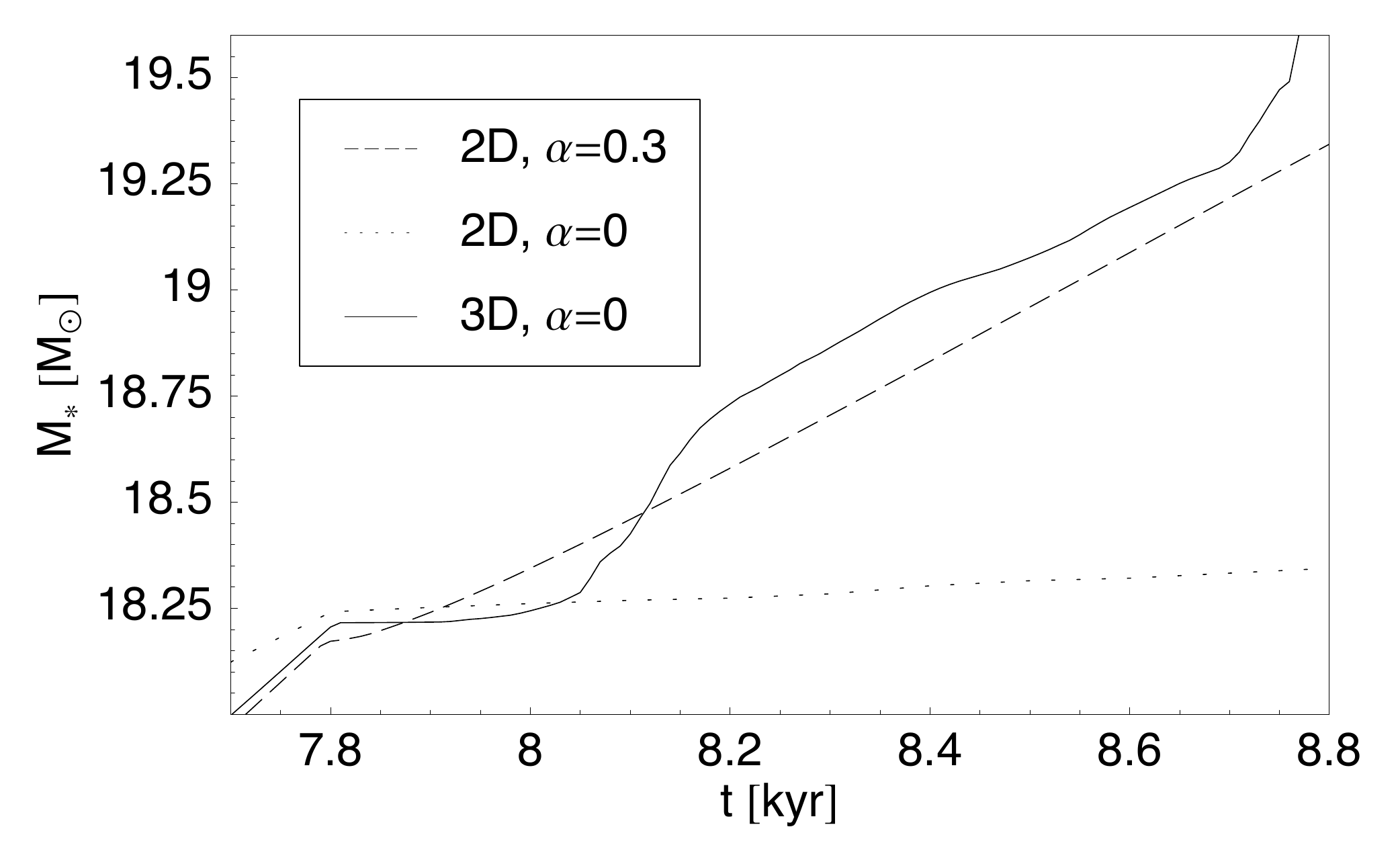}
\caption{
Accretion history of a three-dimensional run, driven by evolving torques in the self-gravitating massive accretion disk, compared with the axially symmetric viscous accretion models with and without $\alpha$-viscosity ($M_\mathrm{core} = 120\Msol$).
}
\label{fig:3D_Accretion}
\end{minipage}
\end{figure}

\newpage
\section{Results}
\label{sect:Results}
The presentation and discussion of the selective simulation results are divided in two subsections, which distinguish between the radiation pressure feedback onto the disk region and its interaction with the accretion flow (Sect.~\ref{sect:Disk}), and the feedback onto the region of the bipolar cavities yielding the launching of stable radiation pressure driven outflows (Sect.~\ref{sect:Outflow}).

\subsection{Radiation pressure in the circumstellar accretion disk}
\label{sect:Disk}
First, we performed pre-stellar core collapse simulations in axial symmetry for four different initial core masses $M_\mathrm{core} = 60 \mbox{ M}_\odot$, $120 \mbox{ M}_\odot$, $240 \mbox{ M}_\odot$, and $480 \mbox{ M}_\odot$.
The resulting accretion histories as a function of the actual stellar mass are displayed in Fig.~\ref{fig:2D_McoreScan}.
Contrary to one-dimensional in-fall simulations 
\citep[e.g.][]{Larson:1971p1210, Kahn:1974p1200, Yorke:1977p1358, Kuiper2010b}, 
the disk accretion scenario allows the formation of massive stars far beyond the upper mass limit of spherically symmetric accretion flows (the so-called `radiation pressure barrier').
The reason for this strong dependence of the radiation pressure feedback on the geometry of the stellar environment is that the massive accretion disk results in a high anisotropy of the radiation field.
Most of the radiation escapes perpendicular to the accretion flow.
Consistently, all simulations performed so far (including resolution studies and different sizes of the inner sink cell) show that the accretion rate increases with a stronger suppression of the isotropic part of the thermal radiation field.

The difference between our simulations and the ones performed by \citet{Yorke:2002p1}, which yield an early and abrupt end of the disk accretion phase directly after the formation of the circumstellar disk, is the fact that in our simulations the radiation dust interaction is resolved down to the dust sublimation front of the forming massive star.
E.g.~in contrast to the size of the inner sink cell of $r_\mathrm{min}=80$~AU or 160~AU as well as a resolution of the inner core region of $(\Delta r)_\mathrm{min} =80$~AU or 160~AU in the 60\Msol or 120\Msol collapse by \citet{Yorke:2002p1} respectively, we use an inner rim of the computational domain of $r_\mathrm{min}=10$~AU with a maximum resolution of $(\Delta r)_\mathrm{min} =1.27$~AU.
For comparison:
Radiation pressure becomes as strong as gravity for roughly a 20\Msol star (the so-called `Generalized Eddington barrier') and its dust sublimation radius in the disk's midplane is about 20 to 30~AU.
The short disk accretion phases by \citet{Yorke:2002p1} are reproduced, if we use the same huge inner sink cell radii they used
\citep[see][]{Kuiper2010b}. 

Secondly, the validity of the standard $\alpha$-shear-viscosity prescription based on \citet{Shakura:1973p3060}, which we use in the axially symmetric disk simulations, is cross-checked in three-dimensional simulations without shear-viscosity, in which the angular momentum transport is self-consistently driven by evolving gravitational torques in the self-gravitating massive accretion disks.
A comparison of the stellar mass growth after the onset of disk formation is shown in Fig.~\ref{fig:3D_Accretion} for the collapse of a 120\Msol pre-stellar core.
Although the three-dimensional run yields a more time dependent episodic accretion history than the viscous disk simulations, the mean accretion rates equal each other.
We conclude that the self-gravity of massive accretion disks is a sufficient driver of angular momentum transport.
The resulting accretion flow easily overcomes the remnant thermal radiation pressure, which is - deeply inside the disk - diminished by the shielding effect at the inner rim of the dust disk.

The difference between our simulations and 
\citet{Krumholz:2007p1380,Krumholz:2009p10975} is the fact that in the disk accretion scenario, which we propose here, the most massive stars can form consistently with standard accretion disk theories without the requirement of `3D radiative Rayleigh-Taylor instabilities' in the outflow region. 
We discuss the issue of the outflow stability in more detail in the following Sect.~\ref{sect:Outflow}.

\subsection{Radiation pressure in the bipolar outflow}
\label{sect:Outflow}
\begin{figure}[tbh]
\centering
\fbox{
\includegraphics[width=0.138\FigureWidth]{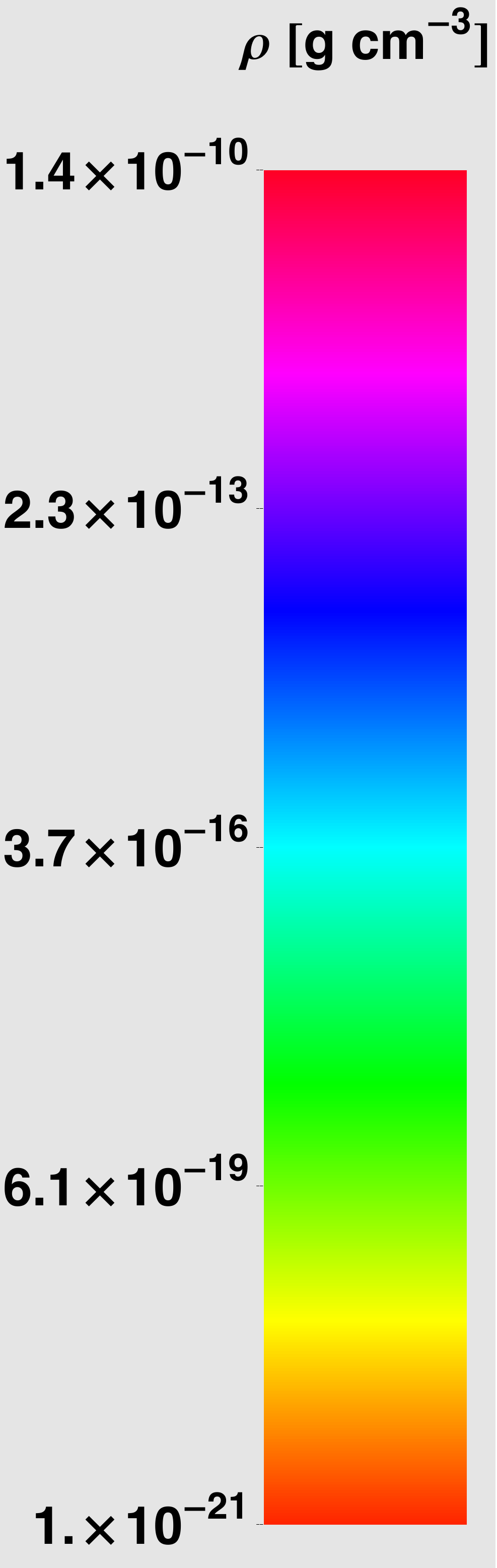}
\hspace{1cm}
\includegraphics[width=0.48\FigureWidth]{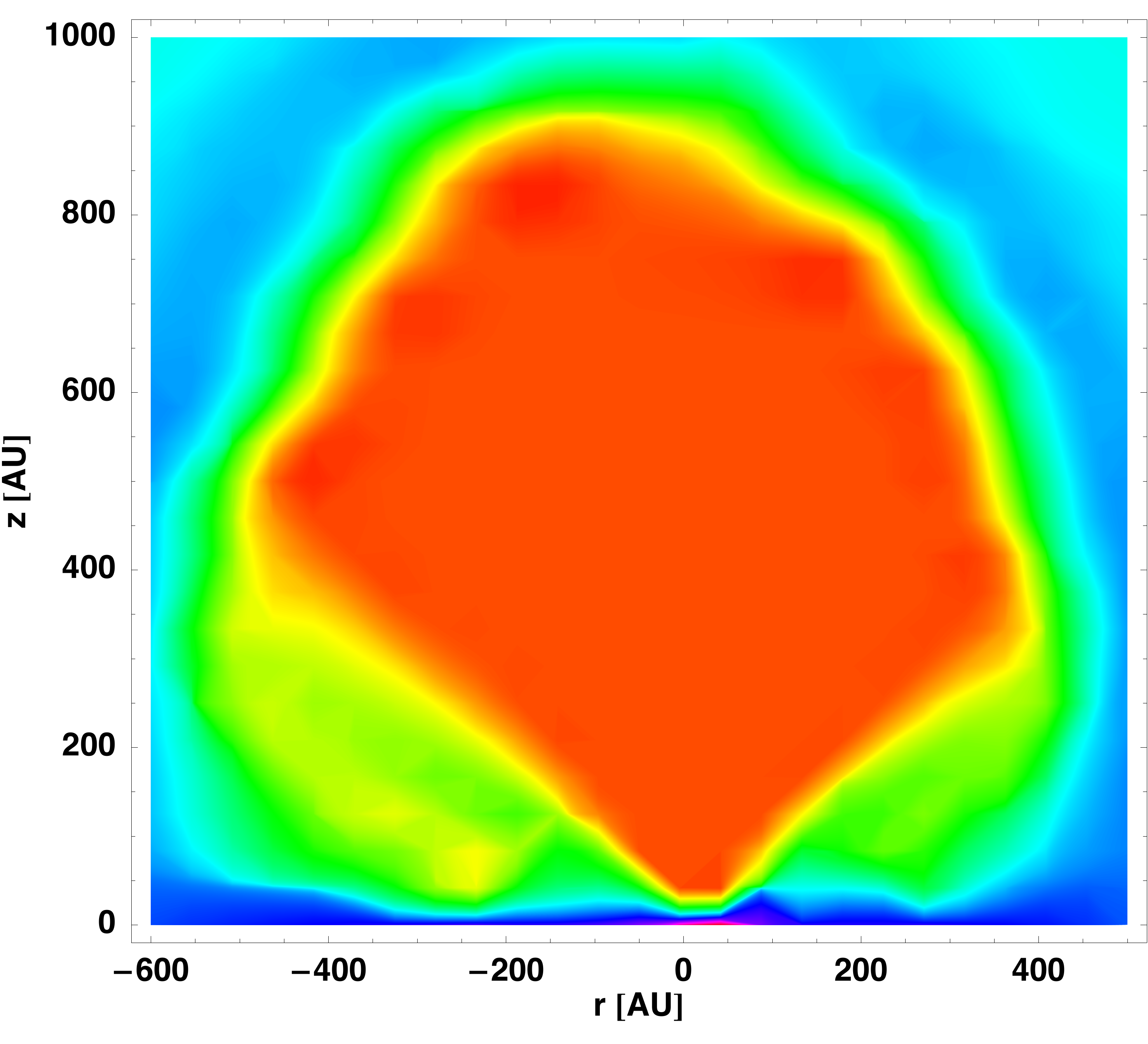}
}
\caption{
A slice through the three-dimensional density structure of a radiation pressure driven outflow during a 120\Msol pre-stellar core collapse.
}
\label{fig:Outflow}
\end{figure}

The feedback of the radiation pressure onto the bipolar region differs strongly from the feedback onto the circumstellar disk region.
The strong radiation pressure in the bipolar direction stops the in-fall and reverts the mass flow into a radiation pressure driven outflow with velocities of a few $100 \mbox{ km s}^{-1}$.
Fig.~\ref{fig:Outflow} visualizes a slice through the three-dimensional density structure of such an outflow.
The radiation pressure driven outflows in our axially symmetric simulations remain stable throughout the whole stellar accretion phase.
As a result, the radiation pressure in the bipolar direction thrusts a substantial fraction (more than $40\%$ during the stellar disk accretion phase) of the initial mass of the pre-stellar core into outer space, reducing the star formation efficiency of the massive star forming core.
In contrast to \citet{Krumholz:2009p10975}, the outflows in our three-dimensional simulations remain stable so far as well in spite of the strong non-axially symmetric features developing in the outflow region.
The reason(s) for that disagreement can be manifold, and we briefly overview the most relevant explanations here:
\begin{list}{\labelenumi}{\leftmargin=1.2em}
\vspace{-1mm}
\addtocounter{enumi}{1}
\item It was supposed by \citet{Krumholz:2009p10975} that the instability requires non-axially symmetric modes, and this could explain why it is not seen in our two-dimensional simulations. 
In our three-dimensional runs, the outflow is stable during the launching and the first growth phase, but instability might appear at a later epoch. 
The instability in \citet{Krumholz:2009p10975} occurs when the outflow extends up to $r \sim 1000$~AU, and the outflow in our runs have propagated up to $r \le 500$~AU so far.
Our three-dimensional simulations are continuously running, and we will further report on their status in the near future.
\vspace{-2mm}
\addtocounter{enumi}{1}
\item Our simulations account for frequency dependent stellar irradiation instead of the frequency averaged (gray) approximation used in \citet{Krumholz:2009p10975}. 
\vspace{-2mm}
\addtocounter{enumi}{1}
\item 
We use a more accurate treatment (ray-tracing) of the first absorption of the stellar irradiation
and the instability could be an artifact of the flux-limited diffusion approximation,
which results in a wrong radiative flux at  $\tau \sim 1$ directly in the cavity wall.
\vspace{-2mm}
\addtocounter{enumi}{1}
\item The disagreement could be due to the different numerical resolution of the diverse grid structures.
\end{list}
Conclusion: The details of the underlying physics of the `3D radiative Rayleigh-Taylor instability' are quite unknown so far and deserve further investigation.

\newpage
\section{Summary}
\label{sect:Summary}
We performed high-resolution radiation hydrodynamics simulations of monolithic pre-stellar core collapses including frequency dependent radiative feedback.
The dust sublimation front of the forming star could be resolved down to 1.27~AU.
The frequency dependent ray-tracing of our newly developed hybrid radiation scheme denotes the most realistic radiation transport method used in multi-dimensional hydrodynamics simulations of massive star formation by now. 
In axial symmetry, the whole accretion phase of several $10^5$~yrs was computed.
The comprehensive parameter studies reveal new insights of the radiative feedback onto the accretion flow during the formation of a massive star:

The massive accretion disk induces a strong anisotropy of the thermal radiation field.
In contrast to the short disk accretion phases in \citet{Yorke:2002p1}, the accretion disks reveal a persistent anisotropy also during their long-term evolution due to the fact that we resolve the dust disk down to the dust sublimation front of the forming star.
Contrary to this explanation, \citet{Krumholz:2009p10975} stated that the much longer accretion phases in their own simulations compared to \citet{Yorke:2002p1} are a result of the so-called `3D radiative Rayleigh-Taylor instability' in the outflow cavity; in axially symmetric simulations sufficient feeding of the disk should therefore be impossible.
But our axially symmetric collapse simulations show a stable radiation pressure driven outflow, while the circumstellar disk gains enough mass from the in-falling envelope to maintain its shielding property over several free fall times, in fact over a longer period ever simulated in previous research studies.

Our three-dimensional core collapse simulations support this accretion disk scenario consistently.
Gravitational torques in the self-gravitating disk lead to a sufficiently high angular momentum transport to maintain accretion.
The accretion history is stronger episodically compared to the viscous disk evolution in the axially symmetric runs, but adds up to a very similar mean accretion rate.

In the bipolar direction, the radiative feedback yields the launch of a stable radiation pressure driven outflow.
No `3D radiative Rayleigh-Taylor instability' is identified yet. 
Although the three-dimensional simulations are still running, we can certainly conclude from the axially symmetric runs that additional feeding by an unstable outflow is not necessary to allow the formation of massive stars.
On the other hand, a well-grounded conclusion on the stability of the outflow and its influence on the formation of massive stars clearly requires further investigation.

Summing up, the central stars in our simulations of the disk accretion scenario grow far beyond the upper mass limit found in the case of spherically symmetric accretion flows. 
For an initial mass of the pre-stellar host core of 60, 120, 240, and 480\Msol the masses of the final stars formed add up to
28.2, 56.5, 92.6, and at least 137.2\Msol respectively.
Indeed, the final massive stars are the most massive stars ever formed in a multi-dimensional radiation hydrodynamics simulation so far. 

%
%

%
%
\footnotesize
%
%
%
%
%
%
%
%
%
%
%
%

\setlength{\bibsep}{0pt}
\bibliographystyle{aa}
\providecommand{\noopsort}[1]{}

\end{document}